\documentclass[runningheads,a4paper]{llncs}
\usepackage{amssymb}
\usepackage{graphicx}
\usepackage[utf8]{inputenc}
\usepackage{url}
\newcommand{\keywords}[1]{\par\addvspace\baselineskip
\noindent\keywordname\enspace\ignorespaces#1}
\usepackage{adjustbox}
\usepackage{algorithm}
\usepackage{algorithmic}
\usepackage{bibnames}
\DeclareUnicodeCharacter{2028}{}

\begin{document}

\mainmatter

\title{Fuzzy Graph Modelling of Anonymous Networks}
\titlerunning{Fuzzy Graph Modelling of Anonymous Networks}

\author{Vasisht Duddu\inst{1} \and Debasis Samanta\inst{2}
\and D. Vijay Rao\inst{3}}
\authorrunning{Duddu et al.}
%

\tocauthor{Vasisht Duddu (Indraprastha Institute of Information Technology),
Debasis Samanta (Indian Institute of Technology),
D. Vijay Rao (Institute for Systems Studies and Analyses )}
\institute{Indraprastha Institute of Information Technology, Delhi, India\\
\email{vduddu@tutamail.com}
\and
Indian Institute of Technology, Kharagpur, India\\
\email{dsamanta@iitkgp.ac.in}
\and
Institute for Systems Studies and Analyses , Delhi, India\\
\email{doctor.rao.cs@gmail.com}}

\toctitle{Fuzzy Graph Modelling of Anonymous Networks}
\tocauthor{Fuzzy Graph Modelling of Anonymous Networks}
\maketitle

\begin{abstract}

Anonymous networks have enabled secure and anonymous communication between the users and service providers while maintaining their anonymity and privacy.
The hidden services in the networks are dynamic and continuously change their domains and service features to maintain anonymity and prevent fingerprinting.
This makes modelling of such networks a challenging task. Further, modelling with crisp graphs is not suitable as they cannot capture the dynamic nature of the anonymous networks.
In this work, we model the anonymous networks using fuzzy graphs and provide a methodology to simulate and analyse an anonymous network.
We consider the case studies of two popular anonymous communication networks: Tor and Freenet, and show how the two networks can be analyzed using our proposed fuzzy representation.

\keywords{Fuzzy Graphs, Anonymous Networks, Network Analysis, Privacy, Anonymity}
\end{abstract}

\section{Introduction}

Communication and social networks have undergone a significant change with the onset of the Internet and rapid technological improvements.
These networks are the backbone of every country's economic and technological growth with large fractions of the businesses moving to Internet based clouds.
Increasing government surveillance and data collection by large corporations, creates the need for privacy and anonymity of the users in the networks.
Recent revelations by Snowden on massive government surveillance has reiterated the importance of private and secure communications \cite{Verble:2014:NES:2684097.2684101}\cite{Greenwald:2014:NPH:2621976}.
Anonymous communication networks are one of the approaches by which users can communicate securely without revealing their identity and send traffic via one or more proxies to obfuscate the true source and destination of messages. These networks use encryption techniques for maintaining the confidentiality and integrity of data.
Some popular anonymous networks are Freenet\cite{freenet}, Tor\cite{Tor}, GNUnet\cite{gnunet} and I2P\cite{i2p}.

Most of the anonymous networks rely on a large number of volunteer run nodes for routing the traffic from the users to the server. These nodes are often unavailable(offline) or malicious, making the network's dependability, reliability, availability  and trust as important parameters to consider in deriving the security metrics.
Recent research in this area shows that in order to mitigate fingerprinting attacks to de-anonymize the hidden services, it is important to have small dynamic websites instead of large, static websites which are more vulnerable to fingerprinting resulting in deanonymizing the services\cite{Overdorf}\cite{Biryukov:2013:TTH:2497621.2498103}.
BuFLO\cite{6234422} proposed a defence against traffic correlation attacks\cite{DBLP:journals/corr/abs-1109-0597} by injecting dummy traffic into the network to obfuscate the genuine traffic which results in varying network communication between the nodes.
To defend against the various attacks and maintain anonymity, a large number of hidden services constantly change their domain names and website content which results in uncertainty in the total available nodes and data communicated for each website in the network.

As a result of the uncertainty in the nodes and traffic in the network, the network cannot be modelled efficiently as a crisp graph as it fails to capture the uncertainity in varying traffic density and unavailability of nodes. In such cases, where some aspects of a graph-theoretic problem are uncertain and vague, it is natural to deal with the uncertainty and vagueness using fuzzy set theory.
We can model the anonymous networks as random, fuzzy graphs, where randomness is associated with either the nodes(fuzzy vertices) or with fuzzy weights(crisp vertices), where the weights denote the strength of association, and other metrics of association between the nodes of the graph.
Modelling of these anonymous overlay networks using fuzzy graphs enables us the following

(i) Better understanding of the underlying dynamics of such networks including routing of data packets which influences the network traffic, delay and congestion.
Such techniques can help mitigate traffic correlation attacks and routing attacks\cite{Johnson:2013:UGR:2508859.2516651}\cite{DBLP:journals/corr/SunEVLRCM15} to deanonymize the users.

(ii) Describing the spatial structure of the networks and predicting changes in the flow pattern due to changes in
the spatial density of nodes and identifying the optimal route to transfer a data.

(iii) Analyse the reliability and error tolerance of anonymous communication networks to DoS attacks and random failures to help build robust networks.

In this work, we make the following main contributions:

- We propose fuzzy graphs to model anonymous communication networks where the nodes and edges of the graphs are modelled as fuzzy variables.

- We provide a methodology to simulate the fuzzy graph model of these networks using percolation algorithm and use network analysis to understand the structure and metrics from the simulation.

- We compare and analyse two popular anonymous communication networks: Tor(an example of Mixnets) and Freenet(an example of Peer to Peer(P2P)) as case studies. We highlight the distinct characteristics of these two networks, to derive their network parameters and structure.

The paper is organised as follows: In Section 2, we describe the related work in this area; fuzzy graph based techniques for representing anonymous networks is introduced in Section 3, and simulating fuzziness in graphs using node percolation algorithm is described in Section 4. This is followed by case studies of TOR and Freenet using the proposed fuzzy representation.

\section{Related Work}

Conventional social networks have been explored extensively and mathematically modelled\cite{Mislove:2007:MAO:1298306.1298311}\cite{article}\cite{1228503} to understand the network properties of modern communication networks.
However, analysing anonymous communication networks is difficult as they use special domains and require additional packages and tools to access them.

A comparison of social networks like Facebook and Twitter with Whisper anonymous network was shown in \cite{Wang:2014:WDA:2663716.2663728} based on network metrics.
Whisper anonymous communication networks has shown to have multiple levels of anonymity for the user's posts and the anonymity of the user's posts vary with demographics like gender, age and education\cite{conf/icwsm/CorreaSMBG15}.
A privacy and network analysis of Tor\cite{Sanchez-Rola:2017:OEC:3038912.3052657} show that the network may leak information by communicating with surface web resulting in de-anonymizing the user.
For understanding the resilience of anonymous networks, network attacks have been shown not only lead to denial of service but also lowering of anonymity as messages need to get retransmitted to be delivered, presenting more opportunities for attack\cite{Borisov:2007:DSD:1315245.1315258}.

However, none of the previous works have mathematically modelled the anonymous communication networks using fuzzy graphs. Most works model the networks as unweighted digraphs(crisp graphs), and do not capture the uncertainty associated with the nodes, vertices, weights, and randomness in the node structures and connectivity, which are the dominant characteristics of anonymous networks.
To address this, we model anonymous networks as fuzzy graphs and propose a method to simulate fuzzy graphs using node percolation.

\section{Fuzzy Graph Representation of Anonymous Networks}

Zadeh introduced a mathematical framework to explain the concept of uncertainty and vagueness in real life in his seminal papers \cite{ZADEH1965338}\cite{Zadeh:1988:FL:44842.44848}\cite{Zadeh:2008:NFL:1365087.1365187}. A fuzzy set is defined mathematically by assigning to each possible individual in the universe of discourse a value, representing its grade of membership, which corresponds to the degree, to which that individual is similar or compatible with the concept represented by the fuzzy set. The fuzzy graph introduced by Rosenfeld \cite{ROSENFELD197577} using fuzzy relations, represents the relationship between the objects by precisely indicating the level of the relationship between the objects of the given set. A number of related concepts based on fuzzy analogous graph theoretic concepts such as bridge, cut vertex and tree were also proposed. Fuzzy graphs have many important applications in modelling real time systems where the level of information inherent in the system varies with different levels of precision\cite{fuzzbook}\cite{Blue:2002:UAF:590141.590149}\cite{YEH1975125}.

\subsection{Fuzzy Graphs: Basic Definitions}

For traditional networks, we model the graphs as an ordered pair $G = (V, E)$ comprising a set $V$ of vertices with a set $E$ of edges or arcs or lines, which are two-element subsets of $V$ comprising of the source and destination.
To model a fuzzy graph, we use the following definitions\cite{survey}:

\paragraph{Definition 1.} A fuzzy subset of a non-empty set $S$ is a mapping $\sigma$ : $S$ $\rightarrow$ [0,1] which assigns to each element $x$ in $S$ a degree of membership, 0 $\leq$ $\sigma(x)$ $\leq$ 1.

\paragraph{Definition 2.} A fuzzy relation on $S$ is a fuzzy subset of $S \times S$. A fuzzy relation $\mu$ on $S$ is a fuzzy relation on the fuzzy subset $\sigma$ if $\mu(x, y)$ $\leq$ $\sigma(x)$ $\wedge$ $\sigma(y)$ for all $x$, $y$ in $S$ where $\wedge$ stands for minimum. A fuzzy relation on the fuzzy subset $\sigma$ is reflexive if $\mu(x, x)$ = $\sigma(x)$ for all $x \epsilon S$. A fuzzy relation $\mu$ on $S$ is said to be symmetric if $\mu(x, y)$ = $\mu(y, x)$ for all $x, y \epsilon S$.

\paragraph{Definition 3.} A fuzzy graph is a pair $G: (\sigma, \mu)$ where $\sigma$ is a fuzzy subset of $S$, $\mu$ is a symmetric fuzzy relation on $\sigma$. The elements of $S$ are called the nodes or vertices of $G$ and the pair of vertices as edges in $G$. The underlying crisp graph of the fuzzy graph $G$ : $(\sigma, \mu)$ is denoted as $G^{*}$ : $(S, E)$ where $E$ $\subseteq$ $S \times S$. The crisp graph $(S, E)$ is a special case of the fuzzy graph $G$ with each vertex and edge of $(S, E)$ having degree of membership 1.

\paragraph{Definition 4.} $(\sigma', \mu')$ is a fuzzy sub graph or a partial fuzzy sub graph of $(\sigma, \mu)$ if $\sigma$' $\subseteq$ $\sigma$ and $\mu$' $\subseteq$ $\mu$; that is if $\sigma$'(u) $\leq$ $\sigma(u)$ for every $u$ $\epsilon$ $S$ and $\mu$'(e) $\leq$ $\mu$(e) for every e $\epsilon$ E.

\paragraph{Definition 5.} $(\sigma', \mu')$  is a fuzzy spanning sub graph of $(\sigma, \mu)$ if $\sigma$' = $\sigma$ and $\mu$' $\subseteq$ $\mu$; that
is if $\sigma$'(u) = $\sigma$(u) for every $u$ $\epsilon$ S and $\mu$'(e) $\leq$ $\mu$(e) for every e $\epsilon$ E. For any fuzzy subset $\nu$ of $S$ such that $\nu$ $\subseteq$ $\sigma$, the fuzzy sub graph of $(\sigma, \mu)$ induced by $n$ is the maximal fuzzy sub graph of $(\sigma, \mu)$, that has fuzzy vertex set $\nu$ and it is the fuzzy sub graph $(\nu, \tau)$ where $\tau(u, v)$ = $\nu(u)$ $\wedge$ $\nu(v)$ $\wedge$ $\mu(u, v)$ for all $u$, $v$ in $S$.

\subsection{Fuzzy Graph Taxonomy} A taxonomy of fuzzy graphs that treats fuzziness in vertex existence, edge existence, edge connectivity, and edge weight was proposed by Bush et.al.\cite{Blue:2002:UAF:590141.590149}. The Types of Fuzzy graphs are classified as follows:

\begin{itemize}
\item \textbf{Type I}: Fuzzy set of Crisp Graphs

\item \textbf{Type II}: Crisp vertex set and Fuzzy edge set: Here, the graph has known vertices but unknown edges. i.e, the vertex set is crisp and the edge set is fuzzy.

\item \textbf{Type III}: Crisp vertices and edges with fuzzy connectivity: In contrast with Type II graph fuzziness, we have a graph with known vertices and edges but unknown edge connectivity, i.e, both the vertex and edge sets are crisp but the edges have fuzzy variables.

\item \textbf{Type IV}: Fuzzy vertex set and Crisp edge set: We have a graph with unknown vertices but known edges, hence, the vertex set is fuzzy and the edge set is crisp.

\item \textbf{Type V}: Crisp graph with fuzzy weights: In this case, the graph has known vertices and edges but unknown weights on the edges, i.e only the weights are fuzzy.

\end{itemize}

In this paper, we use the above taxonomy and definitions to model the anonymous communication networks as fuzzy graphs. The fuzziness in graphs may be modelled by considering (i) the number of vertices(Type IV) to be uncertain or (ii) edge connectivity as random(Type V). In this work, we consider the more common and applicable case of randomness in nodes and simulate the fuzziness in the nodes by proposing random node percolation algorithm for Type IV graphs which is explained in the following section. This approach can be extended to simulate Type V graphs as future work by varying the weights of the edges that represent varying network traffic in the nodes. This study could help explain the dynamics of data flow in the anonymous networks.

\section{Simulating Fuzziness in Graphs using Node Percolation}

In order to capture the uncertainty in the nodes of the anonymous networks, we provide a systematic approach to model and simulate the networks as fuzzy graphs using node percolation algorithm.

\subsection{Anonymous Networks}

Anonymous Networks can be grouped into either mixnets\cite{Chaum:1981:UEM:358549.358563} or P2P\cite{Ratnasamy:2001:SCN:383059.383072} networks based on the underlying protocols. Mixnets use multiple stages of encryption and decryption and the data goes through various permutations to facilitate secure transmission anonymously between client and the server.
On the other hand, P2P networks allow communication among different computers without using a dedicated server to facilitate communication. All the nodes are at the same privilege level and use special protocols for communication within the network. These protocols also include mixnet protocols like onion routing protocols along with public key cryptography for ensuring data integrity and security.
Each node acts as both a supplier and consumer of resources, unlike client-server model, and contributes a part of its computational resources to the network. Routing of traffic is decided using hash tables provided by each node. In this work, we focus on the analysis of two such anonymous networks: Tor and Freenet.
Fig. 1 shows the network structure visualisations of both Tor(Mixnet) and Freenet(P2P) where we observe that Freenet has strongly connected network core while Tor has a densely connected core with a sparse structure along with several node clusters.

\begin{figure}[h]
    \centering
    \includegraphics[width=\textwidth]{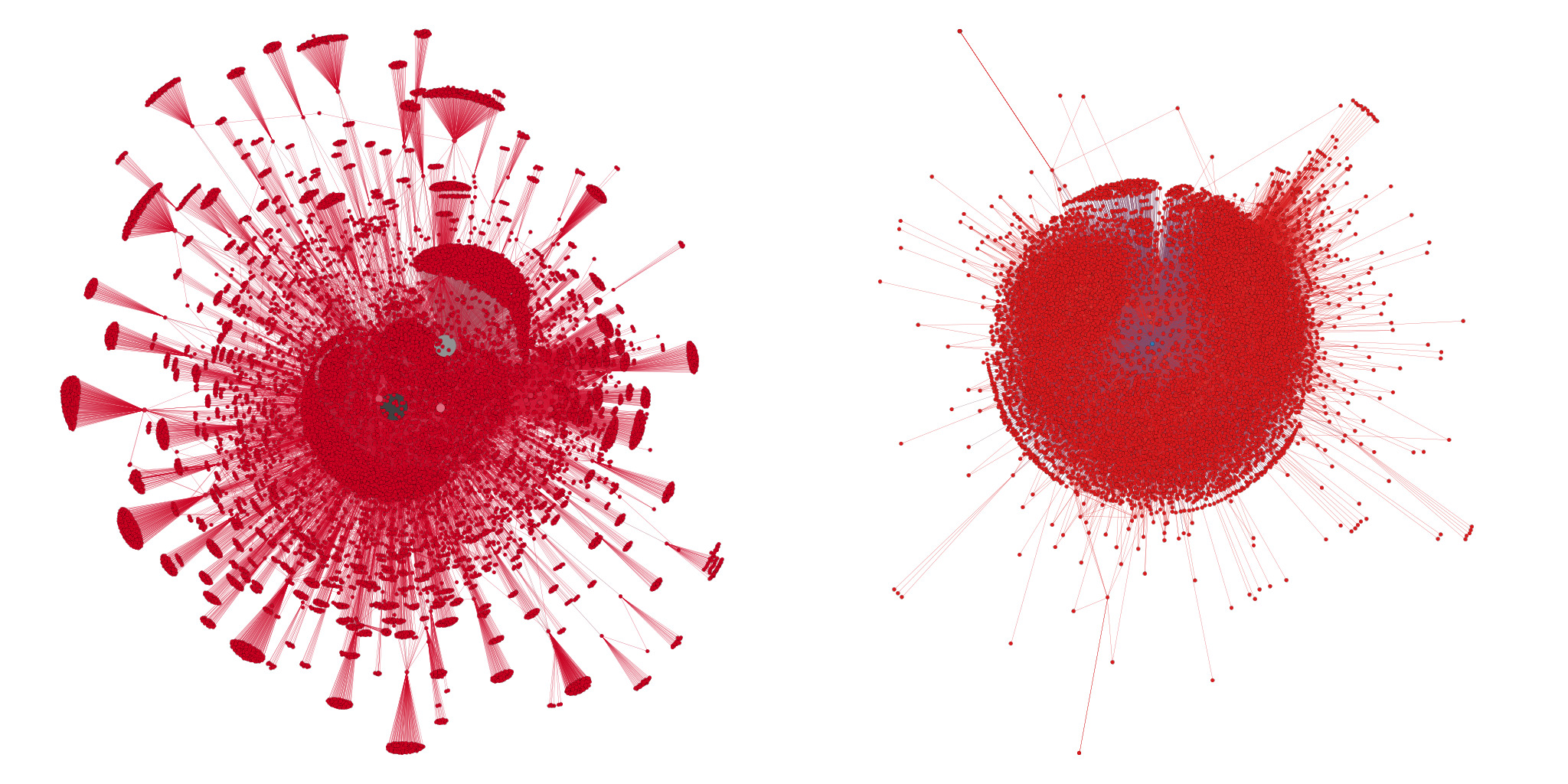}
    \caption{Network visualisation of a) Tor and b) Freenet using Gephi}
\end{figure}

\paragraph{Tor Network.} Tor, acronym for the onion router, is a commonly used anonymous communication network based on Mixnet. It is a second generation low-latency onion routing protocol which is an example of asynchronous free routing mixnet\cite{4077263}. It provides a high throughput with low latency typical for web browsing, emailing, chats and downloading content securely and anonymously and provides perfect forward secrecy\cite{Dingledine:2004:TSO:1251375.1251396}. At each node, a multistage encrypted Tor packet is decrypted to reveal the address of the destination node. Each node only sees the encrypted payload and hence the data remains secure at each node. The server-side anonymity is achieved by using hidden services\cite{hidden}. In case of Tor, onion services are used to host sensitive content such as whistleblower platform, activist blogs and have a strong anonymity requirement.

\paragraph{Freenet Network.} Freenet is based on P2P architecture and is a decentralised network that achieves anonymity by passing requests through multiple intermediary nodes acting as a proxy before reaching the destination node. It operates in two modes: Darknet and Opennet. While Darknet mode of operation allows nodes to connect to known nodes or run by known and trusted people, Opennet allows nodes to connect to any users on the network\cite{freenet}.

\subsection{Data Collection}
For analysing the network structure of both the anonymous communication networks, we crawl through the deep web sites and compile a dataset which includes approximately 15,000 unique nodes representing websites in Tor and different user nodes in case of Freenet. An initial seed list of Tor links was scraped from various online dumps and forums and we used an open source tool\cite{osint}\cite{masc} to index the Tor network and create an edge table from it. For Freenet, data was collected using Web of Trust plugin that created a trust graph of different nodes in the Freenet network\cite{free}. Each dataset had columns containing the source and a destination node which indicated the presence of an edge between them. We filtered out each of the networks to get the giant component and remove isolated nodes. We further set the degree range to ensure relevant nodes are present in the network. We computed network metrics which include: degree distribution, average path length, network diameter, assortativity, network resilience and clustering coefficient using NetworkX library and Gephi.

\subsection{Simulating Fuzziness}

To simulate the uncertainty in the edge connectivity, we randomly remove nodes and corresponding edges using percolation method given in Algorithm 1. We measure the effect of this on the average degree and average path length for different degree of percolation and the results are shown in Fig. 2. In this work, we simulate the fuzziness in the nodes of the network but this can be extended to simulate fuzziness in edge weights by changing the weights and analyse the effect on the networks.

\begin{algorithm}
\begin{algorithmic}[1]

\STATE \textbf{Given:} V=$\{v_1,v_2..,v_n\}$ a set of n vertices in the graph G=(V,E) where E is the set of edges
\STATE \textbf{Input} t: Number of Nodes to remove out of the total nodes
\STATE r $\leftarrow$ 0

\WHILE{$r\not=t$}
  \STATE i $\leftarrow$ rand()  \COMMENT{Generate Index between 1 and n}
  \STATE V.del(i)               \COMMENT{Delete node from vertex set and corresponding edges}
  \STATE $r \leftarrow r + 1$
\ENDWHILE

\RETURN Graph G'=(V',E') where E' is the set of edges after removing $V_t$ vertices and $V'=V-V_t$ is the remaining set of vertices.

\end{algorithmic}
\caption{Random Node Percolation Algorithm to simulate the node fuzziness in anonymous networks}
\end{algorithm}

Generally, removal of a path between two nodes in scale-free networks leads to an increase in the average path length of the network due to increase in distance between any two nodes\cite{article}.
On removing nodes from the network, we observe that the average path length increases for Freenet as expected. Surprisingly, the average path length decreases for Tor network which is unlike typical complex network behaviour. This may be attributed to the fact that the Tor network is sparse and has many isolated node clusters(shown in Section 4.4). As a result of random network attack, these nodes are disconnected from rest of the network resulting in a dense, well-connected core having a lower average path length. In case of average degree, the decrease is gradual in Freenet as compared to Tor. This suggests that the Freenet network is well connected even after the removal of a significant fraction of nodes in the network and indicates that Freenet is more robust to large node failure as compared to Tor. Based on these results, we can clearly distinguish between Mixnets and P2P anonymous networks. P2P networks seem to have an edge over Mixnets in terms of security and robustness of the networks. Freenet being robust to network attacks will be more resistant to Denial of Service attacks as compared to Tor. Mixnets like Tor have a central point of failure while P2P networks have no node hierarchy and hence there is no central point of failure.

\begin{figure}[h]
    \centering
    \includegraphics[width=\textwidth]{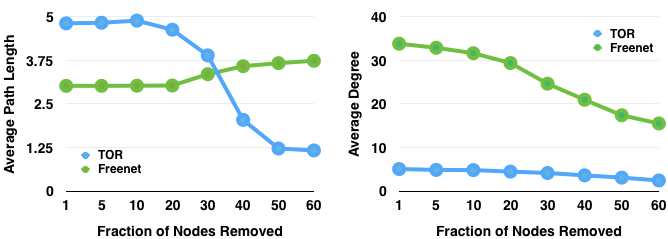}
    \caption{Effect of Random Network Attack on Path Length and Degree}
\end{figure}

These results can be better understood by looking at the network structure of the anonymous communication networks for which we use social network analysis techniques to map the networks and compute the various network metrics.

\subsection{Network Analysis}

\begin{table*}[b!]
  \centering
    \caption{\textbf{Comparing Network Metrics: Tor vs Freenet}}
  \begin{adjustbox}{width=1\textwidth}
  \small
  \begin{tabular}{| c |c | c | c | c | c | c |}
    \hline
    \textbf{Network} & \textbf{Average Degree} & \textbf{Average Path Length} & \textbf{Radius} & \textbf{Diameter} & \textbf{Assortativity} & \textbf{Clustering Coefficient} \\
    \hline
    \textbf{Tor} & 2.982 & 4.630 & 0 & 16 & -0.20  & 0.279\\
    \textbf{Freenet} & 18.362 & 2.965 & 0 & 6 & -0.16 & 0.264\\
    \hline
  \end{tabular}
  \end{adjustbox}

\end{table*}

\paragraph{Degree Distribution.} The degree is defined as the total number of connections of a particular node. We define $p_k$ to be the fraction of vertices in the network that have degree k then a plot of $p_k$ for any given network can be formed by making a histogram of the degrees of vertices which is the degree distribution for the network. A key factor in identifying small world networks is by confirming the scale-free power law distribution. Scale free property of networks is determined by the degree distribution of the nodes. The power law distribution of Tor and Freenet networks are plotted in Fig. 3 which confirms the scale free property since both the networks follow power laws: $p_k$ $\propto$ $k^{- \alpha}$ for some constant exponent $\alpha$.

Table 1 shows the average degree of Freenet to be 18.362 which is quite high as compared to 2.982 for the Tor network. This may be attributed to the P2P architecture of Freenet where no node is privileged over other nodes and most of the nodes actively participate in information storage and retrieval. This results in a very well connected network. Tor, on the other hand, has a large number of nodes connected to a central node which might be acting as a hosting server. The client-server model of Tor results in a lower average degree than Freenet.

\begin{figure}[h]
    \centering
    \includegraphics[width=\textwidth]{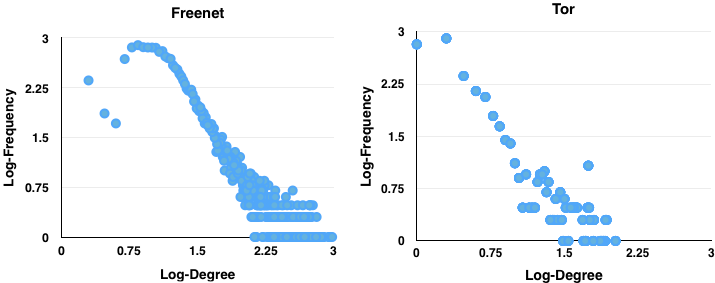}
    \caption{Power Law Distribution between number of nodes and degree on log-log scale}
\end{figure}

\paragraph{Average Path Length.} Average Path Length(L) is defined as the mean distance $d_{ij}$ between two nodes ($i$ and $j$) averaged across all the nodes. Eccentricity is a set of maximum distances between a node and any other node in the network. The diameter of a network is defined as the maximum value in the eccentricity set while the radius of a network is the minimum value of in the eccentricity set.
In the Table, we see that the path length of Tor network is higher as compared to Freenet. The lower value of average path length in Freenet can be due to the high degree of reciprocity among the nodes which are actively engaged in storing and retrieving data. The degree of reciprocity in Tor networks is less due to centralised nature of network resulting in higher average path length. The large value of diameter of Tor as compared to Freenet indicates a sparse Tor network and a dense and compact Freenet network. The radius of both the networks is zero indicating the presence of isolated nodes in the network.

\paragraph{Clustering Coefficient.} Clustering Coefficient(C) is a ratio P/Q, where P is the number of edges between the neighbours of a node and Q is the maximum number of edges that could possibly exist between the neighbours of the node. The value of Clustering Coefficient always lies between 0 and 1 and  Table 1 shows Freenet having a lower number of clusters than Tor network.
Fig. 4 shows the variation of clustering coefficient with respect to out-degree of nodes in both Tor and Freenet and we observe that the nodes with lower out-degree have a higher clustering coefficient in both the networks suggesting that isolated nodes exist in clusters. The nodes in Tor are more concentrated than Freenet for lower outdegree which indicates higher clustering in less connected nodes.
The above results are common characteristic of small world networks.

\begin{figure}[h]
    \centering
    \includegraphics[width=\textwidth]{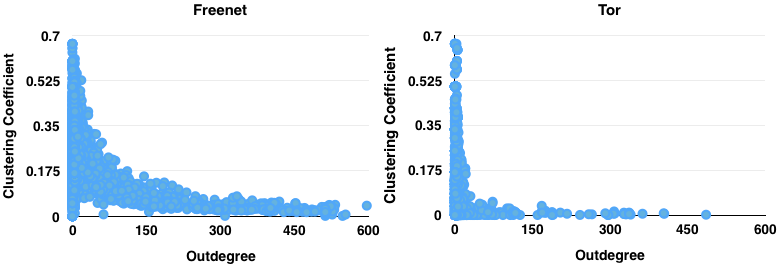}
    \caption{Distribution of Clustering Coefficient with Outdegree}
\end{figure}

\paragraph{Assortativity.} Assortativity is a measure of the probability that nodes of similar degree are attached together and is the Pearson’s correlation coefficient. The range of this coefficient lies between -1 to 1 and a positive or high coefficient suggests that nodes of similar degree are attached to each other while a negative coefficient suggests that the nodes of different degree are connected to each other(preferential attachment).
From Table 1, we see that Tor and Freenet have a high negative assortativity coefficient which suggests that nodes of dissimilar degrees are attached to each other. Tor has a higher negative value as compared to Freenet which indicates that there are a few nodes with a very high degree(hubs or super-hubs) acting as a server or hosting site which are connected to a large number of very low degree nodes which are isolated websites or resources.
In Freenet, the nodes are well connected with a high average degree which explains the higher assortativity coefficient as compared to Tor.

The above analysis using network metrics provides a comprehensive comparison between Tor and Freenet and verifies and explains the results obtained using node percolation algorithm. The proposed systematic methodology to simulate fuzzy graphs can be used to model and analyse various other real world networks like citation network\cite{Tang:08KDD}, temporal networks, location networks and communication networks\cite{snapnets}.

\section{Conclusion}

There is a growing concern of privacy of internet and it is important to analyse the anonymous communication networks which provide the users a means to hide their identity on the internet. We provide a mathematical framework using Fuzzy Graphs to represent the anonymous communication network as a fuzzy graph with fuzzy vertices.
To simulate the fuzziness in the networks, we propose the percolation algorithm to randomly remove nodes and use this approach to analyze two popular anonymous networks: Tor and Freenet.
A comparison between Tor and Freenet using the proposed framework is presented and network analysis and network metrics are derived for the two networks, to understand the structure and effect of random node percolation.

As part of future work, fuzziness in the edges is being studied by using edge percolation algorithm and by changing the weights of the edges. The fuzzy model could be used for modelling dynamic processes within the anonymous networks including traffic flow which can be used for optimising routing algorithms for lower latency.
The proposed methodology helps to analyse, understand the bottlenecks and vulnerabilities in the networks and find mitigations against network attacks in cyber security and cyber warfare scenarios.

\bibliographystyle{splncs03}
\bibliography{sofa.bbl}


\end{document}